\documentclass[prl,twocolumn,superscriptaddress]{revtex4}
\usepackage{amsmath,amssymb,amsfonts,bbm,graphicx}
\usepackage{epstopdf}
\begin{document}
\author{Silvano Garnerone}
\email{garneron@usc.edu}
\affiliation{Department of Physics and Astronomy and Center for Quantum Information Science \& Technology,
 University of Southern California, Los Angeles, CA 90089}
\author{Paolo Giorda}
\affiliation{Institute for Scientific Interchange, Viale Settimio Severo 65, I-10133 Torino, 
Italy}
\author{Paolo Zanardi}
\altaffiliation[Also at ]{Institute for Scientific Interchange, Viale Settimio Severo 65, I-10133 Torino, 
Italy}
\affiliation{Department of Physics and Astronomy and Center for Quantum Information Science \& Technology,
 University of Southern California, Los Angeles, CA 90089}

\date{\today}

\title{Bipartite quantum states and random complex networks}

\begin{abstract}

We introduce  a mapping between  graphs and pure quantum bipartite states
and show that the associated entanglement entropy 
conveys non-trivial information about  the structure of the graph. Our primary goal is to investigate 
the family of random graphs  known as complex networks.
In the case of classical random graphs we derive an analytic expression for the 
averaged entanglement entropy $\bar S$ while for general complex networks we rely on numerics. 
For large number of nodes $n$ we find a scaling $\bar{S} \sim c \log n +g_e$
where both the prefactor $c$ and the  sub-leading $O(1)$ term $g_e$ are a characteristic of
the different classes of complex networks. In particular, $g_e$ encodes topological features
of the graphs and is named network topological entropy.
Our results suggest that quantum entanglement may provide a powerful tool in the analysis
of large complex networks with non-trivial topological properties.
\end{abstract}

\maketitle

\begin{figure}
\includegraphics[scale=0.5]{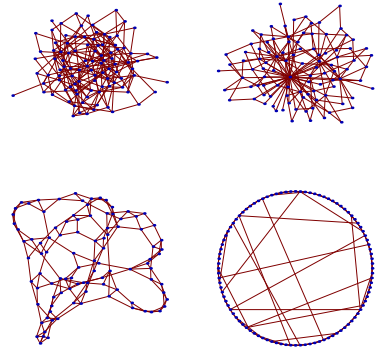} 
\caption{(Color online) Random graphs 
belonging to different ensemble: Erd\H{o}s-R\'{e}nyi (top left corner), 
preferential attachment (top right corner), small-world (bottom left corner), 
and the same graph but with a different embedding (bottom right corner). }
\label{Fig:graphs}
\end{figure}

\textit{Introduction.}
Complex networks are models of graphs 
that appear able to capture the  phenomenology of a plethora of systems, from 
biology to the World Wide Web \cite{book1}. The departure from 
regular lattices, the most common 
background geometry in solid state physics, allows
for the rich static and dynamic behavior 
of complex networks. This is due to the 
simultaneous presence of a global compact structure and 
a sophisticated architecture of interactions. 
By compact structures 
we refer to the typical small  
distance (with respect to a regular lattice) 
between nodes in the network. 
The complexity in the network architecture is 
manifest in the entangled pattern of links and paths 
that such objects display, see Fig.\ref{Fig:graphs}. This architecture encodes 
a type of strong disorder that requires for its 
analysis some of the techniques developed in 
Statistical Mechanics \cite{AlBa}. 
There exist 
a plethora of phenomenological quantities that provide information on the 
architecture of a network: degree distribution, clustering coefficient, 
community structure measures and many others \cite{book1}. 
In particular a few classical entropic measures have been introduced to describe the structure of complex networks \cite{Bia}.\\
\indent
In this letter we address the problem of the entropic analysis and 
discrimination of networks using  quantum information tools, notably entanglement entropy.
Entanglement, a purely quantum measure of correlation, is one of the fundamental  
concepts in quantum information \cite{book2}. 
We provide a recipe, in a way the simplest possible one, to construct a pure bipartite quantum state 
for any given graph.  This allows us to study entanglement properties of quantum states that are 
related to the topological features of the original graphs, and that are able to 
distinguish between different complex network topologies. 
Although at the first sight it may seem a bit artificial  to look for a graph-entropy measure
in a quantum context the synergy between quantum information and complex network tools is not new. 
For example, in \cite{PeLeAc} and  \cite{BrGhSe} the authors have discussed different interesting ways
to associate graphs to quantum states and investigated in which sense complex networks  may play a role in the quantum domain.
All these constructions are then similar in spirit but  substantially different from the present approach.\\
\indent The paper is organized as follows. We first describe the construction 
of the quantum bipartite network states. Then we 
introduce the families of complex networks that we consider in this work. 
Subsequently we define the notion of topological network entropy and 
apply it to study the structure of different complex network topologies. 
Finally we briefly discuss relation with former works and state our conclusions.

\textit{Network quantum states.}
Any graph $ \mathcal{G}_{A} $ with $ n $ nodes 
is completely specified by its 
adjacency matrix $ A $: a 2-dimensional array of size $ n $, 
where each entry $ a_{i,j} $ characterizes the connection 
between nodes $ i $ and $ j $. The domain of $ a_{i,j} $ 
determines the kind of graph one is considering: 
directed, undirected, weighted or unweighted. 
In this work we focus on undirected unweighted graphs, so called simple graphs, 
for which $ a_{i,j} \in \{0,1\}$ (the same analysis can be naturally extended 
to directed and weighted graphs). In particular if 
$ a_{i,j}=0 $ it means there is no edge connecting the two nodes $ i $ and $ j $, 
otherwise $ a_{i,j}=1 $. 
Sometimes it will be useful to refer also to non simple graphs with loops.
Given the graph $ \mathcal{G}_{A} $ ($A\neq 0$) we define the following bipartite 
quantum state 
\begin{equation}
\label{Eq:pure}
|A \rangle \equiv \frac{1 }{\Vert A \Vert_{F}} \sum_{i,j=1}^{n} a_{i,j}|i\rangle |j\rangle 
\in \mathcal{H}_{1}\otimes \mathcal{H}_{2},
\end{equation}
where $ \Vert A \Vert_F:=\sqrt{\mathrm{{Tr}}A^\dagger A}$ denotes the Frobenius norm of the matrix A
and ${\mathcal H}_1\cong{\mathcal H}_2\cong {\mathbb{C}}^n$. 
In the fixed local
basis $ \{ |i\rangle : i=1,\dots,n\} $ we refer to $ |A \rangle $ as a \textit{pure network state}. 
It corresponds to the state of two n-level systems, or analogously to the state of two n-qubit systems where 
each subsystem of n qubits is constrained to the one-excitation manifold.
The isomorphism class of a graph corresponds to the orbit of the permutation group 
on the adjacency matrix: $ PAP^{t}$, where $ P \in S_n $ are permutation matrices.  
This implies that the adjacency matrix of isomorphic graphs is unique up to permutations of rows and columns, and the 
same holds true for the bipartite states in Eq.\ref{Eq:pure}. 
The reduced density matrix of the subsystem whose Hilbert space is $ \mathcal{H}_1 $ 
is given (in the given basis) by 
\begin{equation}
\label{Eq:mixed}
\rho_A \equiv {\mathrm{Tr}}_2 |A \rangle \langle A |
=\frac{AA^{\dagger}}{\|A\|^2_F} \in \mathcal{H}_{1}.
\end{equation}
We refer to this reduced density matrix as the \textit{mixed 
network state}. Notice that both definitions (\ref{Eq:pure}) and (\ref{Eq:mixed})
do not rely on $A$ being symmetric and therefore extends immediately to oriented (and weighted) graphs.
We are interested in the correlation properties  
between $ \mathcal{H}_{1} $ and $ \mathcal{H}_{2} $, 
quantified by the entanglement entropy \cite{PlVi}. 
Since this
quantity depends only on the spectrum of the 
adjacency matrix, it is a property of 
the isomorphism class of the graph, i.e. 
isomorphic graphs will have the same entanglement entropy. 
Indeed, if $P$ is an (orthogonal) permutation matrix: $|PAP^t\rangle= {\hat {P}}^{\otimes\,2}|A\rangle,$
where ${\hat {P}}=\sum_{i,j=1}^n P_{i,j}|i\rangle\langle j|$. Namely isomorphic graphs give rise
to locally equivalent network states.
\\
\indent
Before   considering  complex topologies, it is instructive 
to play with the simplest possible examples and try to characterize 
maximally  and minimally entangled network states.  
Notice that the unnormalized bipartite state can be written as
$ A \otimes I |I \rangle $, where 
$ |I \rangle=\sum_i^n |i,i\rangle $ is an unnormalized maximally entangled 
state. The network corresponding to the state $ |I \rangle $ 
consists of $ n $ nodes with loops, and by construction 
its adjacency matrix is the identity. 
Entanglement does not change under 
local unitary transformations \cite{PlVi}, so in order to construct other 
maximally entangled network states we need to characterize all 
the adjacency matrices that correspond to unitary operators. 
It is easy to prove that the set of unitary adjacency matrices coincides 
with the set of permutation involutions, i.e. the permutation matrices 
that square to the identity. This is also consistent 
with the fact that the square of the adjacency 
matrix is the unnormalized totally mixed network state. 
Unitary adjacency matrices correspond to networks made of only loops or 
disconnected linked pairs of nodes. 
On the other hand factorized states (i.e. unentangled) correspond
to complete graphs with loops.

In the following we study properties of ensembles of random 
network states. The probability measure in the space of network states 
is the one induced by the measure on the space of random networks, 
according to the construction in Eq.\ref{Eq:pure} and Eq.\ref{Eq:mixed}.

\textit{Complex networks.}
In order to make the paper self-contained, let us 
briefly introduce three network structures 
that we will use in the following. 
The seminal paper of Erd\H{o}s and R\'{e}nyi in 1959 
defined what is now the standard example of a random network \cite{ErRe}. 
The ER random graph model, denoted by $ \mathcal{G}^{ER}_{n,m} $, is 
an ensemble of graphs where each element has $ n $ nodes and $ m $ edges. 
ER graphs are also related to so-called 
Gilbert models \cite{Gil}, denoted by $ \mathcal{G}^{ER}_{n,p} $, 
where an edge between each pair of $ n $ nodes 
is present with a probability $ p $. 
The Gilbert model is better suited for analytical investigations, while 
$ \mathcal{G}^{ER}_{n,m} $ graphs are numerically easier to study.
In the thermodynamic limit, fixing the 
average degree $ \overline{q} $ of a node, one can constrain the two 
models to be related by $ \overline{q}=2m/n=pn $.  
If, for $ n\rightarrow \infty $, $ \overline{q}/n \rightarrow 0 $ the network is said to be sparse.
The other example of complex network that we consider 
is known as the Barab\'{a}si-Albert model \cite{BaAl}, denoted by $ \mathcal{G}^{BA} $, 
based on a growth process and a preferential 
attachment mechanism. The rationale is that nodes with higher degree 
acquire new nodes at higher rates than other lower-degree nodes.  
Nodes are added successively, and for each node a number $ d $  of
edges are generated, with bias
towards connections with higher-degree nodes. The distribution for the 
number of links emanating from a node is not 
Poissonian, like for ER graphs, but rather follows
a power-law.
Another way to model stochasticity in the connectivity pattern 
of a graph is by randomly destroying the periodicity 
of a regular lattice. This is the idea behind small-world 
networks, denoted by $ \mathcal{G}^{SW}_{p,k} $, as proposed by Watts and Strogatz \cite{WaSt}. They can 
be created by randomly adding bonds to a regular 
one dimensional ring, this 
way building a superposition between regular lattices 
and classical random graphs. The 
probability $ p $ according to which new bonds are added 
at random is a parameter characterizing the ensemble, 
and it allows to interpolate from regular graphs ($ p=0 $) to 
ER random graphs ($ p=1 $). The other parameter for this 
kind of networks is denoted with $ k $, and it quantifies the number of 
next-nearest-neighbor links present in the original regular graph.
For each of the above complex network ensembles we shall 
construct the associated ensemble of random network states 
denoted by 
$ \psi^{ER}, \psi^{BA} $ and $ \psi^{SW} $, and 
we will consider scaling properties of the 
 average entanglement entropies.

\begin{figure}
\scalebox{0.55}[0.55]{\includegraphics[bb=90 250 1000 550, clip=]{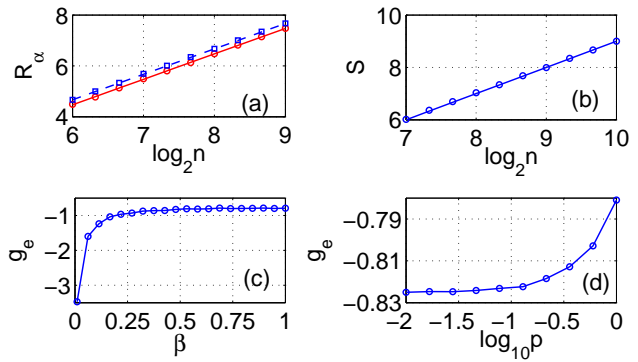}}
\caption{(Color online) (a) The average Renyi entropy $ R_{\alpha} $ obtained numerically, and the analytical 
prediction (dashed and continuous lines) 
for ER quantum random states. The open square symbols corresponds to the $ \alpha=2 $ data points, while the 
open circles are for $ \alpha=3 $.
(b) The average von Neumann entropy obtained numerically and the analytical 
prediction for the states corresponding to $ \mathcal{G}^{ER}_{n,n} $. 
(c) $ g_e $ as a function of $ \beta $ in $ \mathcal{G}^{ER}_{n,\beta n \log_2{n}} $. 
(d) $ g_e $ as a function of the 
rewiring probability $ p $, for small world networks in $ \mathcal{G}^{SW}_{p,2} $.}
\label{Fig:multi}
\end{figure}

\begin{figure}
\includegraphics[scale=0.35]{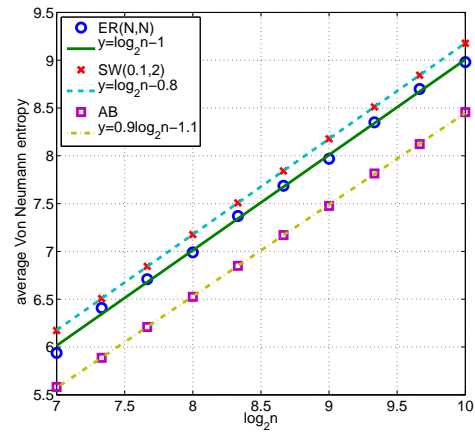} 
\caption{(Color online) The average von Neumann entropy obtained numerically  for different quantum random states 
and the linear fit of the data points. }
\label{Fig:fit3}
\end{figure}

\textit{Topological network entanglement.}
We start by evaluating analytically the averaged Renyi entropy of 
network states in $ \psi^{ER} $, the states 
associated to the ensemble of ER random 
graphs. The $ \alpha $-Renyi entropy 
of a state is defined by
$ R_{\alpha}\left(\rho \right)\equiv (1-\alpha)^{-1}\log_{2}{\rm{Tr}}\rho^{\alpha} $.
Using the definition of $ \rho_A $ given in Eq.\ref{Eq:mixed} 
we have
\begin{equation}
R_{\alpha}\left(\rho_{A} \right)=
\frac{\log_{2}{\rm{Tr}}A^{2\alpha}-\alpha \log_{2}{\rm{Tr}}A^{2}}{1-\alpha}.
\end{equation}
We are interested in the scaling in $ n $ of the average Renyi 
entropy $ \overline{R_{\alpha}} $. In order to provide an explicit expression 
for $ \overline{R_{\alpha}} $ we use the fact that for each $ \alpha $ there exists 
a constant $ c_{2\alpha} $ such that
$ \lim_{n\rightarrow \infty} n^{-1}\overline{{\rm{Tr}}A^{2\alpha}}=c_{2\alpha}   $,
i.e. for sparse ER graphs the thermodynamic limit of the moments of the graph spectrum exist and 
are finite \cite{Spi}. 
Furthermore one can check numerically that 
the difference between the quenched average $ (1-\alpha)^{-1}\overline{\log_2 \rho^{\alpha}} $ and the 
annealed average $ (1-\alpha)^{-1}\log_2 \overline{\rho^{\alpha}} $ scales like $ n^{-1} $.
Putting this together we can write
\begin{equation}
\label{Eq:renyER}
\overline{R_{\alpha}\left(\rho_{A} \right)}\left[n\right]=
\log_{2}n+ g(\alpha)+O(\frac{1}{n} ).
\end{equation}
where $g(\alpha):=\log_2 c_2+(1-\alpha)^{-1}(\log_2 c_{2\alpha}-\log_2 c_{2})$
is a sub-leading $O(1)$ term.
This equation tells us that the Renyi entropy is almost maximal for any $ \alpha $.
Notice that, even though the logarithmic scaling for these network states is consistent with the one of general 
(Haar distributed) random states \cite{Pag}, one could not predict a priori this behavior for the particular 
family of random states we introduced. 
Remarckably the sub-leading term contains information about the topology of the graphs. 
In fact, the term $ c_{2\alpha} $ is directly related to the average number 
of closed paths of length $ 2\alpha $ in the graph. 
In Fig.\ref{Fig:multi}a we provide a numerical check of Eq.\ref{Eq:renyER}, which 
supports in particular the approximation of the quenched with the annealed average. 
The figure shows a perfect agreement between Eq.\ref{Eq:renyER} and the results of the simulation.
It is tempting to extrapolate our analysis from the Renyi entropy to 
the von Neumann entanglement entropy, which is defined by $\overline{S}\equiv\lim_{\alpha \rightarrow 1} 
\overline{R_{\alpha}}$.
By construction it follows immediately the logarithmic scaling
of the von Neumann entropy, while for the sub-leading term we have 
$g_e:=\lim_{\alpha\to 1} g(\alpha)=\log_2 c_2- \frac{d\log c_{2x}}{dx}|_{x=1}$
and then, using the definition of $c_{2\alpha}$, one finds
\begin{equation}
g_e= \log_2{c_2}-\frac{\overline{{\rm{Tr}}A^2\log_2{A^2}}}{\overline{{\rm{Tr}}A^2}}.
\label{Eq:ge}
\end{equation}
It is interesting to observe that the second term in Eq.\ref{Eq:ge} can be regarded as a sort of topological susceptibility
of the given family of networks. In fact this term is equal to $- \frac{\log c_{2x}}{dx}|_{x=1}$, and it tells us 
how the logarithm of the rescaled averaged number of loops of length $2\alpha$ changes 
as the length is changed continuously around $\alpha=1.$ For this reason, and in view of  its conceptual similarity
with the topological entanglement entropy introduced in \cite{KiPr} we call the $O(1)$ quantity in Eq.\ref{Eq:ge} 
\textit{topological network entanglement}. 
In Fig.\ref{Fig:multi}b it is shown a comparison between the analytical expression for the 
von Neumann entropy, obtained using Eq.\ref{Eq:ge}, and the empirical 
average of the entanglement entropy over different realizations. 
As can be seen from the figure there is a perfect overlap between the two. 
The ensemble $ \mathcal{G}^{ER}_{n,m} $ is parameterized by the number of edges $ m $. 
One can wonder about the behavior of $ g_e $ as $ m $ varies in some particular interval, but 
in such a way that the graph is always sparse. 
Fig. \ref{Fig:multi}c shows the value of $ g_e $ as a function of a parameter
$ \beta \in (0,1]$ which is related to the number of edges by $m=\beta n \log_2{n} $. 
The figure shows that for ER-graphs the greater the number of links the greater the entanglement. 
We now evaluate numerically $ g_e $ for small worlds and preferential attachment networks.
Considering first networks in $ \mathcal{G}^{SW}_{p,k} $ we checked numerically that 
the scaling is logarithmic and that the prefactor is always $ 1 $. 
In Fig.\ref{Fig:multi}d we see the dependence of $ g_e $ on $ p $, 
the probability of rewiring edges. The figure shows that $ g_e $ 
increases monotonically from regular to more random graphs. This is 
consistent with the intuition that adding randomness to a graph 
increases its entropy, as measured by $ g_e $. From these results 
it is clear that the properties of the entanglement entropy 
provide information on the complex network structure, supporting 
its interpretation as a graph entropic quantity. 
Considering the ability to discriminate between different network topologies, in
Fig.\ref{Fig:fit3} we compare 
the scaling of the average von Neumann entropy of $ \psi^{ER}_{n,n} $, $ \psi^{SW}_{0.1,2} $, and $ \psi^{BA} $
random network states. The simulations show that, unlike for ER and small-world networks, the logarithmic 
prefactor for $ \psi^{BA} $ states is slightly smaller than $ 1 $.  
From the figure it is clear that the von Neumann entropy
distinguishes different complex network ensembles. For sufficiently big networks 
the fluctuations due to disorder are strongly suppressed. On the one hand this is 
an indication of the robustnes of this graph-entropic measure; on the other hand 
it can also be useful from a computational point of view. In fact one has 
a very good estimate of the entanglement entropy already from few realizations.
Hence, if the network is big enough the scaling analysis could in principle 
be done on one single realization, for example evaluating the entanglement entropy on 
sub-graphs of increasing size. Furthermore whenever computational efficiency 
is an issue we point out that instead of the Von Neumann entropy one could 
evaluate the so called single-copy entanglement ($\lim_{\alpha \rightarrow \infty} R_{\alpha}$) \cite{EiCr}, 
for which efficient numerical techniques can be used \cite{Saa}.

\textit{Discussion and conclusions.}
In this letter we have exploited a natural mapping from graphs to quantum bipartite states and we have defined 
the entropy of a graph as the entanglement entropy of the associated quantum state.
We have then used this quantum measure of correlations to study the structure of complex networks.  The scaling of the entanglement entropy is 
logarithmic in the system dimension, and  both the prefactor and the sub-leading $O(1)$ term
(topological network entropy)  can be used to characterize the network family and to distinguish between different network topologies. 
In particular, we showed that the Barab\'{a}si-Albert model has a scaling 
behavior that differs significantly from the one of small-world and ER graphs. 
While these last two have a similar scaling, but still distinguishable 
comparing graphs with the same number of edges. This is consistent 
with the fact that small-world networks are mixture of regular lattices 
and ER graphs. Furthermore, we provided an analytic expression, exact in the 
thermodynamic limit, for the averaged Renyi and von Neumann entropy associated to ER random graphs.
It is desirable to achieve a clear and general understanding  of 
the relations between the quantum entropic measures 
we introduced and the standard 
graph-theoretic observables analyzed in the complex-network community.  One would like also to gain a deeper insight into the measure concentration (large size convergence) properties of the various probabilistic objects we discussed for the different families of complex networks. 
While the primary goal of this paper has been to show how  to use quantum tools to investigate complex networks it should be clear that also the converse task i.e., using properties of complex newtwork to study the novel class of random quantum states we introduced, is of interest on its own right. Moreover 
on the quantum side it is a challenge to find a consistent inverse mapping that allows one to  associate to a general bipartite quantum state a specific network. Finally, one would like to devise efficient and physically feasible preparation schemes for the network quantum states we proposed.
\\
\indent
We thank N. Toby Jacobson for a careful reading of the manuscript. P.Z. acknowledges
support from NSF grants PHY-803304, DMR-0804914

\end{document}